\documentclass[a4paper,11pt]{article}

\makeatletter

\gdef\@fpheader{\newline}
\gdef\@journal{jhep}

\RequirePackage{amsmath}
\RequirePackage{amssymb}
\RequirePackage{epsfig}
\RequirePackage{CJK}
\RequirePackage{graphicx}
\RequirePackage[numbers,sort&compress]{natbib}
\RequirePackage{color}
\RequirePackage[colorlinks=true
,urlcolor=blue
,anchorcolor=blue
,citecolor=blue
,filecolor=blue
,linkcolor=blue
,menucolor=blue
,pagecolor=blue
,linktocpage=true
,pdfproducer=medialab
,pdfa=true
,CJKbookmarks
]{hyperref}

\newif\ifnotoc\notocfalse
\newif\ifemailadd\emailaddfalse
\newif\iftoccontinuous\toccontinuousfalse

\def\@subheader{\@empty}
\def\@keywords{\@empty}
\def\@abstract{\@empty}
\def\@xtum{\@empty}
\def\@dedicated{\@empty}
\def\@arxivnumber{\@empty}
\def\@collaboration{\@empty}
\def\@collaborationImg{\@empty}
\def\@proceeding{\@empty}
\def\@preprint{\@empty}

\newcommand{\subheader}[1]{\gdef\@subheader{#1}}
\newcommand{\keywords}[1]{\if!\@keywords!\gdef\@keywords{#1}\else%
\PackageWarningNoLine{\jname}{Keywords already defined.\MessageBreak Ignoring last definition.}\fi}
\renewcommand{\abstract}[1]{\gdef\@abstract{#1}}
\newcommand{\dedicated}[1]{\gdef\@dedicated{#1}}
\newcommand{\arxivnumber}[1]{\gdef\@arxivnumber{#1}}
\newcommand{\proceeding}[1]{\gdef\@proceeding{#1}}
\newcommand{\xtumfont}[1]{\textsc{#1}}
\newcommand{\correctionref}[3]{\gdef\@xtum{\xtumfont{#1} \href{#2}{#3}}}
\newcommand\jname{JHEP}
\newcommand\acknowledgments{\section*{Acknowledgments}}

\newcommand\preprint[1]{\gdef\@preprint{\hfill #1}}

\newcommand\note[2][]{%
\if!#1!%
\stepcounter{footnote}\footnotetext{#2}%
\else%
{\renewcommand\thefootnote{#1}%
\footnotetext{#2}}%
\fi}

\newtoks\auth@toks
\renewcommand{\author}[2][]{%
  \if!#1!%
    \auth@toks=\expandafter{\the\auth@toks#2\ }%
  \else
    \auth@toks=\expandafter{\the\auth@toks#2$^{#1}$\ }%
  \fi
}

\newtoks\affil@toks\newif\ifaffil\affilfalse
\newcommand{\affiliation}[2][]{%
\affiltrue
  \if!#1!%
    \affil@toks=\expandafter{\the\affil@toks{\item[]#2}}%
  \else
    \affil@toks=\expandafter{\the\affil@toks{\item[$^{#1}$]#2}}%
  \fi
}

\newtoks\email@toks\newcounter{email@counter}%
\newcommand{\emailAdd}[1]{%
\emailaddtrue%
\ifnum\theemail@counter>0\email@toks=\expandafter{\the\email@toks, \@email{#1}}%
\else\email@toks=\expandafter{\the\email@toks\@email{#1}}%
\fi\stepcounter{email@counter}}
\newcommand{\@email}[1]{\href{mailto:#1}{\tt #1}}

\newcommand*\collaboration[1]{\gdef\@collaboration{#1}}
\newcommand*\collaborationImg[2][]{\gdef\@collaborationImg{#2}}

\newcommand\afterLogoSpace{\smallskip}
\newcommand\afterSubheaderSpace{\vskip3pt plus 2pt minus 1pt}
\newcommand\afterProceedingsSpace{\vskip21pt plus0.4fil minus15pt}
\newcommand\afterTitleSpace{\vskip23pt plus0.06fil minus13pt}
\newcommand\afterRuleSpace{\vskip23pt plus0.06fil minus13pt}
\newcommand\afterCollaborationSpace{\vskip3pt plus 2pt minus 1pt}
\newcommand\afterCollaborationImgSpace{\vskip3pt plus 2pt minus 1pt}
\newcommand\afterAuthorSpace{\vskip5pt plus4pt minus4pt}
\newcommand\afterAffiliationSpace{\vskip3pt plus3pt}
\newcommand\afterEmailSpace{\vskip16pt plus9pt minus10pt\filbreak}
\newcommand\afterXtumSpace{\par\bigskip}
\newcommand\afterAbstractSpace{\vskip16pt plus9pt minus13pt}
\newcommand\afterKeywordsSpace{\vskip16pt plus9pt minus13pt}
\newcommand\afterArxivSpace{\vskip3pt plus0.01fil minus10pt}
\newcommand\afterDedicatedSpace{\vskip0pt plus0.01fil}
\newcommand\afterTocSpace{\bigskip\medskip}
\newcommand\afterTocRuleSpace{\bigskip\bigskip}
\newlength{\affiliationsSep}
\newcommand\beforetochook{\pagestyle{myplain}\pagenumbering{roman}}

\DeclareFixedFont\trfont{OT1}{phv}{b}{sc}{11}

\renewcommand\maketitle{
\pagestyle{empty}
\thispagestyle{titlepage}
\setcounter{page}{0}
\noindent{\small\scshape\@fpheader}\@preprint\par
\afterLogoSpace
\if!\@subheader!\else\noindent{\trfont{\@subheader}}\fi
\afterSubheaderSpace
\if!\@proceeding!\else\noindent{\sc\@proceeding}\fi
\afterProceedingsSpace
{\LARGE\flushleft\sffamily\bfseries\@title\par}
\afterTitleSpace
\hrule height 1.5\p@%
\afterRuleSpace
\if!\@collaboration!\else
{\Large\bfseries\sffamily\raggedright\@collaboration}\par
\afterCollaborationSpace
\fi
\if!\@collaborationImg!\else
{\normalsize\bfseries\sffamily\raggedright\@collaborationImg}\par
\afterCollaborationImgSpace
\fi
{\bfseries\raggedright\sffamily\the\auth@toks\par}
\afterAuthorSpace
\ifaffil\begin{list}{}{%
\setlength{\leftmargin}{0.28cm}%
\setlength{\labelsep}{0pt}%
\setlength{\itemsep}{\affiliationsSep}%
\setlength{\topsep}{-\parskip}}
\itshape\small%
\the\affil@toks
\end{list}\fi
\afterAffiliationSpace
\ifemailadd %% if emailadd is true
\noindent\hspace{0.28cm}\begin{minipage}[l]{.9\textwidth}
\begin{flushleft}
\textit{E-mail:} \the\email@toks
\end{flushleft}
\end{minipage}
\else %% if emailaddfalse do nothing
\PackageWarningNoLine{\jname}{E-mails are missing.\MessageBreak Plese use \protect\emailAdd\space macro to provide e-mails.}
\fi
\afterEmailSpace
\if!\@xtum!\else\noindent{\@xtum}\afterXtumSpace\fi
\if!\@abstract!\else\noindent{\renewcommand\baselinestretch{.9}\textsc{Abstract:}}\ \@abstract\afterAbstractSpace\fi
\if!\@keywords!\else\noindent{\textsc{Keywords:}} \@keywords\afterKeywordsSpace\fi
\if!\@arxivnumber!\else\noindent{\textsc{ArXiv ePrint:}} \href{http://arxiv.org/abs/\@arxivnumber}{\@arxivnumber}\afterArxivSpace\fi
\if!\@dedicated!\else\vbox{\small\it\raggedleft\@dedicated}\afterDedicatedSpace\fi
\ifnotoc\else
\iftoccontinuous\else\newpage\fi
\beforetochook\hrule
\tableofcontents
\afterTocSpace
\hrule
\afterTocRuleSpace
\fi
\setcounter{footnote}{0}
\pagestyle{myplain}\pagenumbering{arabic}
} % close the \renewcommand\maketitle{

\renewcommand{\baselinestretch}{1.1}\normalsize
\divide\@tempdima\baselineskip
\@tempcnta=\@tempdima
\skip\@mpfootins = \skip\footins
\renewcommand{\@dotsep}{10000}

\newcommand\ps@myplain{
\pagenumbering{arabic}
\renewcommand\@oddfoot{\hfill-- \thepage\ --\hfill}
\renewcommand\@oddhead{}}
\let\ps@plain=\ps@myplain

\newcommand\ps@titlepage{\renewcommand\@oddfoot{}\renewcommand\@oddhead{}}

\numberwithin{equation}{section}

\renewcommand\section{\@startsection{section}{1}{\z@}%
                                   {-3.5ex \@plus -1.3ex \@minus -.7ex}%
                                   {2.3ex \@plus.4ex \@minus .4ex}%
                                   {\normalfont\large\bfseries}}
\renewcommand\subsection{\@startsection{subsection}{2}{\z@}%
                                   {-2.3ex\@plus -1ex \@minus -.5ex}%
                                   {1.2ex \@plus .3ex \@minus .3ex}%
                                   {\normalfont\normalsize\bfseries}}
\renewcommand\subsubsection{\@startsection{subsubsection}{3}{\z@}%
                                   {-2.3ex\@plus -1ex \@minus -.5ex}%
                                   {1ex \@plus .2ex \@minus .2ex}%
                                   {\normalfont\normalsize\bfseries}}
\renewcommand\paragraph{\@startsection{paragraph}{4}{\z@}%
                                   {1.75ex \@plus1ex \@minus.2ex}%
                                   {-1em}%
                                   {\normalfont\normalsize\bfseries}}
\renewcommand\subparagraph{\@startsection{subparagraph}{5}{\parindent}%
                                   {1.75ex \@plus1ex \@minus .2ex}%
                                   {-1em}%
                                   {\normalfont\normalsize\bfseries}}

\def\fnum@figure{\textbf{\figurename\nobreakspace\thefigure}}
\def\fnum@table{\textbf{\tablename\nobreakspace\thetable}}

\long\def\@makecaption#1#2{%
  \vskip\abovecaptionskip
  \sbox\@tempboxa{\small #1. #2}%
  \ifdim \wd\@tempboxa >\hsize
    \small #1. #2\par
  \else
    \global \@minipagefalse
    \hb@xt@\hsize{\hfil\box\@tempboxa\hfil}%
  \fi
  \vskip\belowcaptionskip}

\renewenvironment{thebibliography}[1]{%
\begin{oldthebibliography}{#1}%
\small%
\raggedright%
\setlength{\itemsep}{5pt plus 0.2ex minus 0.05ex}%
}%
{%
\end{oldthebibliography}%
}

\makeatother

\usepackage[T1]{fontenc} % if needed
\usepackage{romannum}

\begin{document} %正文开始

%%%%%%%%题目作者%%%%%%%%%%%%%%%%%%%%%%%%%%%%%%%%%%%%%%%%%%%%%%%%

\title{{\boldmath Quantum correction of gravitational constant} \\  }
%% %simple case: 2 authors, same institution
%% \author{A. Uthor}
%% \author{and A. Nother Author}
%% \affiliation{Institution,\\Address, Country}
\author[a]{Yu-Jie Chen,}
\author[a]{Shi-Lin Li,}
\author[b]{Yu-Zhu Chen,}\note{chenyuzhu@nankai.edu.cn,}
\author[c]{Wen-Du Li,}\note{liwendu@tjnu.edu.cn,}
\author[a]{and Wu-Sheng Dai}\note{daiwusheng@tju.edu.cn.}

\affiliation[a]{Department of Physics, Tianjin University, Tianjin 300350, P.R. China}
\affiliation[b]{Theoretical Physics Division, Chern Institute of Mathematics, Nankai University, Tianjin, 300071, P. R. China}
\affiliation[c]{College of Physics and Materials Science, Tianjin Normal University, Tianjin 300387, PR China}

% more complex case: 4 authors, 3 institutions, 2 footnotes
%\author[a,b,1]{author1,\note{Corresponding author.}}
%\author[b]{author2}
%\author[a,2]{T. Hird\note{Also at Some University.}}
%\author[a,2]{and Fourth}

% The "\note" macro will give a warning: "Ignoring empty anchor..."
% you can safely ignore it.

%\affiliation[a]{Department of Physics, Tianjin University,\\Tianjin 300072, P.R. China}
%\affiliation[b]{Another University,different-address, Country}
%\affiliation[c]{A School for Advanced Studies,\\some-location, Country}

% e-mail addresses: one for each author, in the same order as the authors
%\emailAdd{first@tju.edu.cn}
%\emailAdd{second@tju.edu.cn}
%\emailAdd{third@one.univ}
%\emailAdd{fourth@one.univ}

%%%%%%%%%%%%%%%%%%%%%%%%%%%%%%%%%%%%%%%%%%%%%%%%%%%%%%%%%%%%%%%%

%%%%%%%%%%%%%摘要和关键字%%%%%%%%%%%%%%%%%%%%%%%%%%%%%%%%%%%%%%%

\abstract{
We suggest a scheme for considering the quantum correction of the
gravitational constant. In the model, the gravitational constant originates
from a coupling of the gravitational field with a scalar field. In this paper,
we show that if the scalar field, as it should be in the real physical world,
is a quantum field, then the gravitational constant will have a
spacetime-dependent quantum correction, so that the quantum corrected physical
constant is no longer a constant. The quantum correction of the gravitational
constant is different in different spacetime. We calculate the quantum
correction in the Schwarzschild spacetime, the $H_{3}$ (Euclidean $AdS_{3}$)
spacetime, the $H_{3}/Z$ spacetime, the universe model, the de Sitter
spacetime, and the Rindler spacetime.
}

%\keywords{......}

%%%%%%%%%%%%%%%%%%%%%%%%%%%%%%%%%%%%%%%%%%%%%%%%%%%%%%%%%%%%%%%%

%%%%%%%%%%%%%%%正文%%%%%%%%%%%%%%%%%%%%%%%%%%%%%%%%%%%%%%%%%%%%%

\maketitle %生成题目

\flushbottom
%(正文开始) ——————————————————————————————————————————————————

\section{Introduction}

The gravitational constant is determined phenomenologically. There is no
generally accepted theory of the origin of the gravitational constant. In the
frame of the induced gravity model, the gravitational constant originates from
a coupling of the gravitational field and a scalar field. In the classical
picture of the induced gravity, the scalar field is a classical field and the
gravitational constant is a constant equaling to the gravitational constant in
the Einstein gravity. However, in the real world, a scalar field, as it should
be, must be a quantum field rather than a classical field. We show that if the
scalar field is a quantum field, there will be a quantum correction on the
gravitational constant. The quantum corrected gravitational constant is no
longer a constant at the quantum level, but varies with time and space.

The quantum correction of the gravitational constant is different in different
spacetimes. In this paper, as examples, we calculate the quantum correction of
the gravitational constant in the Schwarzschild spacetime, the $H_{3}$
(Euclidean $AdS_{3}$) spacetime, the $H_{3}/Z$ spacetime, the universe models,
the de Sitter spacetime, and the Rindler spacetime.

In this paper, we show that the gravitational constant may have a quantum
correction. This correction is small, only at a quantum level. Nevertheless,
there are studies show that even a small change in the gravitational constant
will lead a remarkable influence on the evolution of the universe, for big
bang nucleosynthesis is very sensitive to the value of the gravitational
constant \cite{alvey2020improved}. This inspires us to study the influence of
such a quantum level change of the gravitational constant in cosmology.

In this paper, the gravitational constant has a time- and space-dependent
quantum correction. The result of the present paper provides a physical
mechanism of nonconstant gravitational constants. The variation of
gravitational constants with time has been discussed for a long time
\cite{dirac1937cosmological,teller1948change}. The effect of a nonconstant
gravitational constant in cosmology is also considered theoretically and
experimentally \cite{alvey2020improved,braglia2020larger,nagata2004wmap}. In
cosmology indued gravity theory has been applied, such as Bianchi-I
cosmological models \cite{kamenshchik2018induced}, the reconstruction
technique of the potential for scalar fields in induced-gravity based
cosmological models \cite{kamenshchik2011reconstruction}, induced gravity dark
energy models \cite{ballardini2016cosmological}, the quantum backreaction of
the scalar and the tensor perturbations in induced gravity
\cite{tronconi2011quantum}, and the exorcism of ghosts in induced gravity
\cite{narain2017exorcising}. Black holes \cite{nandan2010black} and black hole
entropies \cite{frolov2003cft} in induced gravity are considered. Induced
gravity with scalar fields and non-Abelian gauge interactions are studied
\cite{einhorn2016inducedI,einhorn2016induced}. The Schwinger-DeWitt technique
for the covariant curvature expansion of effective actions for brane induced
gravity are developed \cite{barvinsky2010schwinger,ravanpak2019dynamical}. The
criterion defining single-field models leading to Starobinsky-like inflation
\cite{giudice2014starobinsky-like}, the induced gravity on the
Dvali-Gabadadze-Porrati brane
\cite{nozari2012stability,nozari2012braneworld,bouhmadilopez2011cosmology},
and the Wheeler-DeWitt equation for induced gravity models
\cite{kamenshchik2019induced} are studied.

In section \ref{Gravity}, we consider the quantum correction of the
gravitational constant. As examples, we calculate the quantum correction of
the gravitational constant in various spacetimes, including the Schwarzschild
spacetime, the $H_{3}$ (Euclidean $AdS_{3}$) spacetime, the $H_{3}/Z$
spacetime, the universe model, the de Sitter spacetime, and the Rindler
spacetime. Section \ref{Conclusion} is devoted to conclusions and outlook.

\section{Quantum correction of gravitational constant \label{Gravity}}

The action of gravity is%
\begin{equation}
I=\int d^{4}x\sqrt{-g}\frac{1}{16\pi G}R, \label{Gaction}%
\end{equation}
where $G$ is the gravitational constant and $R$ is the Ricci scalar. First
construct an induced gravity model. Introduce a scalar field $\Phi$ and couple
the scalar field $\Phi$ to the gravitational field:%
\begin{equation}
I=\int d^{4}x\sqrt{-g}\left[  \chi\left(  \Phi\right)  R+\frac{1}{2}g^{\mu\nu
}\partial_{\mu}\Phi\partial_{\nu}\Phi-V\left(  \Phi\right)  \right]  .
\label{2.1.0}%
\end{equation}
Zee's induced gravity \cite{zee1979broken} can be recovered by taking
$\chi\left(  \Phi\right)  =\frac{1}{2}\epsilon\Phi^{2}$. Comparison with Eq.
(\ref{Gaction}) shows that $\chi\left(  \Phi\right)  $ plays the role of the
reciprocal of the gravitational constant $G$.

The equation of the scalar field $\Phi$, by the action (\ref{2.1.0}), reads%
\begin{equation}
\frac{1}{\sqrt{-g}}\partial_{\mu}\left(  \sqrt{-g}g^{\mu\nu}\partial_{\nu}%
\Phi\right)  =-\frac{dV_{\text{eff}}\left(  \Phi\right)  }{d\Phi},
\label{Phieq}%
\end{equation}
where%
\begin{equation}
V_{\text{eff}}\left(  \Phi\right)  =-\chi\left(  \Phi\right)  R+V\left(
\Phi\right)  . \label{2.1.6}%
\end{equation}

In the model, the gravitational constant $G$ is the reciprocal of $\chi\left(
\Phi\right)  $. In the classical picture, the gravitational constant $G$ is a
constant, so $\chi\left(  \Phi\right)  $ must be a constant. This requires
that the solution of the field $\Phi$ must be a stable classical ground-state
solution: $\Phi=\Phi_{0}$, so that $\chi\left(  \Phi\right)  =\chi\left(
\Phi_{0}\right)  $ is a constant. The classical ground state $\Phi=\Phi_{0}$
is the position of the minimum value of $V_{\text{eff}}\left(  \Phi\right)  $
given by $V_{\text{eff}}^{\prime}\left(  \Phi_{0}\right)  =0$. In the
classical picture, the gravitational constant is a constant and is independent
of the gravitational field itself. This requires that though in Eq.
(\ref{2.1.6}) $V_{\text{eff}}\left(  \Phi\right)  $ depends on the Ricci
scalar $R$, the position of the minimum value of $V_{\text{eff}}\left(
\Phi\right)  $ should not depend on $R$. For this purpose, we choose the
position of the minimum value of $\chi\left(  \Phi\right)  $ and the position
of the minimum value of $V\left(  \Phi\right)  $ are the same, so that the
position of the minimum value of $V_{\text{eff}}\left(  \Phi\right)  $ is the
same as the minimum value of $\chi\left(  \Phi\right)  $ and the minimum value
of $V\left(  \Phi\right)  $. Consequently, the classical ground state
$\Phi=\Phi_{0}$ which is the position of the minimum value of $V_{\text{eff}%
}\left(  \Phi\right)  $ is also the position of the minimum values of
$\chi\left(  \Phi\right)  $ and $V\left(  \Phi\right)  $, i.e., $V_{\text{eff}%
}^{\prime}\left(  \Phi_{0}\right)  =V^{\prime}\left(  \Phi_{0}\right)
=\chi^{\prime}\left(  \Phi_{0}\right)  =0$.

In the classical picture, the gravitational constant is
\begin{equation}
G=\frac{1}{16\pi\chi\left(  \Phi_{0}\right)  }. \label{2.1.9}%
\end{equation}
The scalar field $\Phi$ is in the classical ground state: $\Phi=\Phi_{0}$. The
classical ground state $\Phi=\Phi_{0}$ is the position of the minimum values
of $\chi\left(  \Phi\right)  $ and $V\left(  \Phi\right)  $ and the minimum
values of $\chi\left(  \Phi\right)  $ and $V\left(  \Phi\right)  $ are the same.

Nevertheless, in the real physical world a scalar field must be a quantum
field. We will show that a quantum scalar field leads to a spacetime-dependent
quantum corrected gravitational constant.

Next we construct the equation of quantum correction.

The action of a scalar field in the spacetime with the metric $g_{\mu\nu}$ is%
\begin{equation}
S\left[  \Phi\right]  =\int\sqrt{-g}d^{4}x\mathcal{L}\left(  \Phi\right)  .
\end{equation}
The classical field $\Phi_{\text{cl}}$ is given by the extreme value of the
action functional:
\begin{equation}
\left.  \frac{\delta S\left[  \Phi\right]  }{\delta\Phi}\right\vert
_{\Phi=\Phi_{\text{cl}}}=0,\label{0-1}%
\end{equation}
i.e., $\Phi_{\text{cl}}$ is the solution of the classical field equation
(\ref{0-1}); note that the classical field equation with the action
(\ref{2.1.0}) is Eq. (\ref{Phieq}). Expanding the action $S\left[
\phi\right]  $ around the classical field $\phi_{\text{cl}}$, keeping the
expansion to the second order, and applying the classical field equation
(\ref{0-1}) give%
\begin{align}
S\left[  \Phi\right]   &  =S\left[  \Phi_{\text{cl}}\right]  \nonumber\\
&  +\frac{1}{2}\int\sqrt{-g\left(  x_{1}\right)  }d^{4}x_{1}\sqrt
{-g_{2}\left(  x_{2}\right)  }d^{4}x_{2}\left.  \frac{\delta^{2}S\left[
\Phi\right]  }{\delta\Phi\left(  x_{1}\right)  \delta\Phi\left(  x_{2}\right)
}\right\vert _{\phi=\phi_{\text{cl}}}\left[  \Phi\left(  x_{1}\right)
-\Phi_{\text{cl}}\left(  x_{1}\right)  \right]  \left[  \Phi\left(
x_{2}\right)  -\Phi_{\text{cl}}\left(  x_{2}\right)  \right]  .\label{0-4}%
\end{align}

The generating functional $Z\left[  J\right]  =\int\mathcal{D}\Phi\exp\left(
\frac{i}{\hbar}\left(  S\left[  \Phi\right]  +\int\sqrt{-g}d^{4}xJ\Phi\right)
\right)  $ with the action (\ref{0-4}), by using the Gaussian integral formula
$\int\mathcal{D}\Phi\exp\left(  -\left[  \frac{1}{2}\left(  \Phi,A\Phi\right)
+\left(  b,\Phi\right)  +c\right]  \right)  =\left(  \det A\right)
^{-1/2}\exp\left[  \frac{1}{2}\left(  b,A^{-1}b\right)  -c\right]  $, reads%
\begin{align}
Z\left[  J\right]   &  =e^{\frac{i}{\hbar}S\left[  \Phi_{\text{cl}}\right]
+\frac{i}{\hbar}\int\sqrt{-g\left(  x\right)  }d^{4}xJ\left(  x\right)
\Phi_{\text{cl}}\left(  x\right)  }\left[  \det\left(  -\frac{i}{\hbar}\left.
\frac{\delta^{2}S\left[  \Phi\right]  }{\delta\Phi\left(  x_{1}\right)
\delta\Phi\left(  x_{2}\right)  }\right\vert _{\Phi=\Phi_{\text{cl}}}\right)
\right]  ^{-1/2}\nonumber\\
&  \times\exp\left(  -\frac{i}{2\hbar}\left[  \int\sqrt{-g\left(
x_{1}\right)  }d^{4}x_{1}\sqrt{-g_{2}\left(  x_{2}\right)  }d^{4}x_{2}J\left(
x_{1}\right)  \left(  \left.  \frac{\delta^{2}S\left[  \Phi\right]  }%
{\delta\Phi\left(  x_{1}\right)  \delta\Phi\left(  x_{2}\right)  }\right\vert
_{\Phi=\Phi_{\text{cl}}}\right)  ^{-1}J\left(  x_{2}\right)  \right]  \right)
.\label{0-8}%
\end{align}

The generating functional $W\left[  J\right]  $, defined by $Z\left[
J\right]  =e^{\frac{i}{\hbar}W\left[  J\right]  }$, by Eq. (\ref{0-8}), reads%
\begin{align}
W\left[  J\right]   &  =-i\hbar\ln Z\left[  J\right]  \nonumber\\
&  =S\left[  \Phi_{\text{cl}}\right]  +\int d^{4}x\sqrt{-g\left(  x\right)
}J\left(  x\right)  \Phi_{\text{cl}}\left(  x\right)  +\frac{i}{2}%
\hbar\operatorname{tr}\left(  \ln\left(  -\frac{i}{\hbar}\left.  \frac
{\delta^{2}S\left[  \Phi\right]  }{\delta\Phi\left(  x_{1}\right)  \delta
\Phi\left(  x_{2}\right)  }\right\vert _{\Phi=\Phi_{\text{cl}}}\right)
\right)  \nonumber\\
&  -\frac{1}{2}\int\sqrt{-g\left(  x_{1}\right)  }d^{4}x_{1}\sqrt
{-g_{2}\left(  x_{2}\right)  }d^{4}x_{2}J\left(  x_{1}\right)  \left(  \left.
\frac{\delta^{2}S\left[  \Phi\right]  }{\delta\Phi\left(  x_{1}\right)
\delta\Phi\left(  x_{2}\right)  }\right\vert _{\phi=\phi_{\text{cl}}}\right)
^{-1}J\left(  x_{2}\right)  .\label{09}%
\end{align}

The effective action can be then obtained by the Legendre transform
$\Gamma\left[  \Phi\right]  =W\left[  J\right]  -\int\sqrt{-g\left(  x\right)
}d^{4}x\Phi\left(  x\right)  J\left(  x\right)  $:%
\begin{align}
\Gamma\left[  \Phi\right]   &  =S\left[  \Phi_{\text{cl}}\right]  +\frac{i}%
{2}\hbar\operatorname{tr}\left[  \ln\left(  -\frac{i}{\hbar}\left.
\frac{\delta^{2}S\left[  \Phi\right]  }{\delta\Phi\left(  x_{1}\right)
\delta\Phi\left(  x_{2}\right)  }\right\vert _{\Phi=\Phi_{\text{cl}}}\right)
\right]  \nonumber\\
&  -\int\sqrt{-g\left(  x_{1}\right)  }d^{4}x_{1}\sqrt{-g_{2}\left(
x_{2}\right)  }d^{4}x_{2}\frac{3}{2}\left.  \frac{\delta^{2}S\left[
\Phi\right]  }{\delta\Phi\left(  x_{1}\right)  \delta\Phi\left(  x_{2}\right)
}\right\vert _{\phi=\phi_{\text{cl}}}\eta\left(  x_{1}\right)  \eta\left(
x_{2}\right)  .\label{14}%
\end{align}
The extreme value of the effective action functional, by $\frac{\delta
\Gamma\left[  \Phi\right]  }{\delta\Phi}=0$, gives the equation of the quantum
fluctuation $\eta$:
\begin{equation}
\int\sqrt{-g\left(  x_{1}\right)  }d^{4}x_{1}\left(  \left.  \frac{\delta
^{2}S\left[  \Phi\right]  }{\delta\Phi\left(  x_{1}\right)  \delta\Phi\left(
x_{2}\right)  }\right\vert _{\phi=\phi_{\text{cl}}}\right)  \eta\left(
x_{1}\right)  =0.\label{15}%
\end{equation}

This method is equivalent to writing%
\begin{equation}
\Phi=\Phi_{\text{cl}}+\eta, \label{Phiqc}%
\end{equation}
and directly substituting Eq. (\ref{Phiqc}) into the classical equation
(\ref{Phieq}), as that in Ref. \cite{graham2009spectral}.

The equation of the quantum correction, with the action (\ref{2.1.0}), up to
the first-order contribution reads:%
\begin{equation}
\frac{1}{\sqrt{-g}}\partial_{\mu}\left(  \sqrt{-g}g^{\mu\nu}\partial_{\nu}%
\eta\right)  +\left[  V^{\prime\prime}\left(  \Phi_{\text{cl}}\right)
-\chi^{\prime\prime}\left(  \Phi_{\text{cl}}\right)  R\right]  \eta=0.
\label{2.2.6}%
\end{equation}

Expanding $\chi\left(  \Phi\right)  $ around $\eta=0$, we have%
\begin{equation}
\chi\left(  \Phi\right)  =\chi\left(  \Phi_{\text{cl}}+\eta\right)
=\chi\left(  \Phi_{\text{cl}}\right)  +\frac{1}{2}\chi^{\prime\prime}\left(
\Phi_{\text{cl}}\right)  \eta^{2}+\cdots. \label{XiPhi}%
\end{equation}
The reason why the leading contribution is the $\eta^{2}$ term is that the
$\eta$ term vanishes for $\chi^{\prime}\left(  \Phi_{\text{cl}}\right)  =0$.
The quantum corrected gravitational constant by Eqs. (\ref{2.1.9}) and
(\ref{XiPhi}) then reads%
\begin{equation}
G_{Q}=\frac{1}{16\pi\chi\left(  \Phi\right)  }\simeq\frac{1}{16\pi\chi\left(
\Phi_{\text{cl}}\right)  }\left[  1-\frac{1}{2}\frac{\chi^{\prime\prime
}\left(  \Phi_{\text{cl}}\right)  }{\chi\left(  \Phi_{\text{cl}}\right)  }%
\eta^{2}\right]  . \label{GQ}%
\end{equation}
The quantum correction $\eta$ is given by Eq. (\ref{2.2.6}). In our case, the
classical solution of the scalar field is the classical ground-state solution,
i.e., $\Phi_{\text{cl}}=\Phi_{0}$, so%
\begin{equation}
G_{Q}=G\left[  1-\frac{1}{2}\frac{\chi^{\prime\prime}\left(  \Phi_{0}\right)
}{\chi\left(  \Phi_{0}\right)  }\eta^{2}\right]
\end{equation}
with $G$ the Newtonian gravitational constant. The quantum corrected
gravitational constant depends on the spacetime metric $g^{\mu\nu}$. In the
classical picture, the gravitational constant in different spacetime is the
same, but the quantum corrected gravitational constant in different spacetimes
are different.

In the following, we will not use the concrete function form of $\chi\left(
\Phi\right)  $ and $V\left(  \Phi\right)  $. The result apples to any
$\chi\left(  \Phi\right)  $ and $V\left(  \Phi\right)  $ so long as they are
lower bounded. In different theories $\chi\left(  \Phi\right)  $ and $V\left(
\Phi\right)  $ are different, e.g., in Zee's induced gravity $\chi\left(
\Phi\right)  =\frac{1}{2}\epsilon\Phi^{2}$ \cite{zee1979broken} and in
$\Phi^{4}$-theory $V\left(  \Phi\right)  =\lambda\Phi^{4}$.

The equation of the quantum fluctuation $\eta$, Eq. (\ref{2.2.6}), is a linear
homogeneous equation, so $\eta$ and $C\eta$ are both the solutions of Eq.
(\ref{2.2.6}). Therefore the quantum fluctuation $\eta$ can only be determined
up to a constant factor. In other words, there is an undetermined constant in
the quantum fluctuation $\eta$.

In physical experiments, at a spacetime point $x_{0}=\left(  t_{0}%
,\mathbf{x}_{0}\right)  $, one can measure the experimental value of the
gravitational constant $G^{\text{experiment}}\left(  x_{0}\right)  $. For
example, in\ the universe model one can choose $t_{0}$ as the current time and
in Schwarzschild spacetime one can choose $x_{0}$ as the spacetime point where
the observer stands. At $x_{0}$, the quantum corrected gravitational constant
by Eqs. (\ref{2.1.9}) and (\ref{GQ}) is%
\begin{equation}
G_{Q}\left(  x_{0}\right)  =\frac{1}{16\pi\chi\left(  \Phi_{\text{cl}}\left(
x_{0}\right)  \right)  }\left[  1-\frac{1}{2}\frac{\chi^{\prime\prime}\left(
\Phi_{\text{cl}}\left(  x_{0}\right)  \right)  }{\chi\left(  \Phi_{\text{cl}%
}\left(  x_{0}\right)  \right)  }\eta^{2}\left(  x_{0}\right)  \right]  .
\label{C1}%
\end{equation}
The classical solution $\Phi_{\text{cl}}\left(  x\right)  $ is given by
solving the usual classical scalar field equation. Then we have%
\begin{equation}
G_{Q}\left(  x_{0}\right)  =G^{\text{experiment}}\left(  x_{0}\right)  .
\label{GeqG}%
\end{equation}
The quantum fluctuation at $x_{0}$ then by Eq. (\ref{GeqG}) reads
\begin{equation}
\eta\left(  x_{0}\right)  =\sqrt{2\frac{\chi\left(  \Phi_{\text{cl}}\left(
x_{0}\right)  \right)  }{\chi^{\prime\prime}\left(  \Phi_{\text{cl}}\left(
x_{0}\right)  \right)  }\left[  1-16\pi\chi\left(  \Phi_{\text{cl}}\left(
x_{0}\right)  \right)  G_{Q}^{\text{experiment}}\left(  x_{0}\right)  \right]
}. \label{Constant}%
\end{equation}
The undetermined constant in the quantum fluctuation $\eta$ can be solved from
Eq. (\ref{Constant}).

In our model, as discussed above, the world is in its ground state, so the
classical solution is the classical ground-state solution: $\Phi_{\text{cl}%
}\left(  x_{0}\right)  =\Phi_{0}$. Then Eq. (\ref{Constant}) becomes
$\eta\left(  x_{0}\right)  =\sqrt{2\frac{\chi\left(  \Phi_{0}\right)  }%
{\chi^{\prime\prime}\left(  \Phi_{0}\right)  }\left[  1-16\pi\chi\left(
\Phi_{0}\right)  G_{Q}^{\text{experiment}}\left(  x_{0}\right)  \right]  }$.

\section{Schwarzschild spacetime}

Consider the quantum correction of the gravitational constant in the
Schwarzschild spacetime.

In the Schwarzschild spacetime,%
\begin{equation}
ds^{2}=-\left(  1-\frac{2M}{r}\right)  dt^{2}+\frac{1}{1-\frac{2M}{r}}%
dr^{2}+r^{2}d\theta^{2}+r^{2}\sin^{2}\theta d\varphi^{2},
\end{equation}
the radial equation of the field, by Eq. (\ref{Phieq}), after variable
separation $\Phi\left(  t,r,\theta,\varphi\right)  =\cos\left(  \omega
t\right)  \phi\left(  r\right)  Y\left(  \theta,\varphi\right)  $, reads%
\begin{equation}
\left(  1-\frac{2M}{r}\right)  \phi^{\prime\prime}\left(  r\right)
+\frac{2\left(  r-M\right)  }{r^{2}}\phi^{\prime}\left(  r\right)
-\frac{l\left(  l+1\right)  }{r^{2}}\phi\left(  r\right)  +V^{\prime}\left(
\phi\left(  r\right)  \right)  +\frac{\omega^{2}}{1-\frac{2M}{r}}\phi\left(
r\right)  =0,
\end{equation}
where we assume that the potential depends only on the radial part of the
field, $V\left(  \Phi\right)  =V\left(  \phi\left(  r\right)  \right)  $ .

In the classical picture, the fact that the gravitational constant given by
Eq. (\ref{2.1.9}) is a constant requires that the classical solution is a
stable classical ground-state solution. Therefore, we only concern ourselves
with the classical ground state which corresponds to $\omega=0$ and $l=0$:%
\begin{equation}
\left(  1-\frac{2M}{r}\right)  \phi^{\prime\prime}\left(  r\right)
+\frac{2\left(  r-M\right)  }{r^{2}}\phi^{\prime}\left(  r\right)  +V^{\prime
}\left(  \phi\left(  r\right)  \right)  =0. \label{5.1.3}%
\end{equation}

The equation of $\eta$ is given by substituting $\phi=\phi_{\text{cl}}+\eta$
into Eq. (\ref{5.1.3}) and keeping only the leading contribution:%
\begin{equation}
\frac{r-2M}{r}\eta^{\prime\prime}\left(  r\right)  +\frac{2\left(  r-M\right)
}{r^{2}}\eta^{\prime}\left(  r\right)  +V^{\prime\prime}\left(  \phi
_{\text{cl}}\right)  \eta\left(  r\right)  =0. \label{5.1.4}%
\end{equation}

To solve Eq. (\ref{5.1.4}), we use the variable substitution $z=\frac{r}{M}-1$
to convert Eq. (\ref{5.1.4}) to a confluent Heun equation
\cite{li2019scattering}:
\begin{equation}
\frac{d}{dz}\left[  \left(  z^{2}-1\right)  \frac{dy\left(  z\right)  }%
{dz}\right]  +\left[  -p^{2}\left(  z^{2}-1\right)  +2p\beta z-\lambda
-\frac{m^{2}+s^{2}+2msz}{z^{2}-1}\right]  y\left(  z\right)  =0, \label{CHeun}%
\end{equation}
where $p=M\sqrt{-V^{\prime\prime}\left(  \phi_{\text{cl}}\right)  }$, $m=s=0$,
$\beta=-M\sqrt{-V^{\prime\prime}\left(  \phi_{\text{cl}}\right)  }$, and
$\lambda=-2V^{\prime\prime}\left(  \phi_{\text{cl}}\right)  M^{2}$. The
solution of the confluent Heun equation (\ref{CHeun}) is the confluent Heun
function $y\left(  z\right)  =H_{c}^{\left(  \alpha\right)  }\left(
p,\alpha,\gamma,\delta,\sigma;z\right)  $ with $\gamma=m+s+1$, $\delta=m-s+1$,
$\alpha=-\beta+m+1$, and $\sigma=\lambda-2p\left(  \beta-m-s-1\right)
-m\left(  m+1\right)  $ \cite{ronveaux1995heun}. Then the quantum fluctuation
reads%
\begin{equation}
\eta=CH_{c}^{\left(  \alpha\right)  }\left(  M\sqrt{-V^{\prime\prime}\left(
\phi_{\text{cl}}\right)  },\left[  M\sqrt{-V^{\prime\prime}\left(
\phi_{\text{cl}}\right)  }+1\right]  ,1,1,\right.  \left.  \left[
-4M^{2}V^{\prime\prime}\left(  \phi_{\text{cl}}\right)  +2M\sqrt
{-V^{\prime\prime}\left(  \phi_{\text{cl}}\right)  }\right]  ;\frac{r}%
{M}-1\right)  .
\end{equation}

The quantum corrected gravitational constant by Eq. (\ref{GQ}) is%
\begin{align}
G_{Q}  &  =\frac{1}{16\pi\chi\left(  \phi_{\text{cl}}\right)  }\left\{
1-\frac{C^{2}}{2}\frac{\chi^{\prime\prime}\left(  \Phi_{\text{cl}}\right)
}{\chi\left(  \Phi_{\text{cl}}\right)  }\right.  \left[  H_{c}^{\left(
\alpha\right)  }\left(  M\sqrt{-V^{\prime\prime}\left(  \phi_{\text{cl}%
}\right)  },\left[  M\sqrt{-V^{\prime\prime}\left(  \phi_{\text{cl}}\right)
}+1\right]  ,1,1,\right.  \right. \nonumber\\
&  \left.  \left.  \left.  \left[  -4M^{2}V^{\prime\prime}\left(
\phi_{\text{cl}}\right)  +2M\sqrt{-V^{\prime\prime}\left(  \phi_{\text{cl}%
}\right)  }\right]  ;\frac{r}{M}-1\right)  \right]  ^{2}\right\}  .
\label{GQS}%
\end{align}
The gravitational constant in the Schwarzschild spacetime depends on the
radial coordinate $r$.

The constant $C$ is determined by observation. Suppose the observer stands at
$r_{0}$. The gravitational constant measured by an observer standing at
$r_{0}$ is $G^{\text{experiment}}\left(  r_{0}\right)  $. Then $G_{Q}\left(
r_{0}\right)  =G^{\text{experiment}}\left(  r_{0}\right)  $ determines the
constant $C$:%
\begin{align}
C  &  =\sqrt{\frac{2\chi\left(  \phi_{0}\right)  }{\chi^{\prime\prime}\left(
\phi_{0}\right)  }\left[  1-16\pi\chi\left(  \phi_{0}\right)
G^{\text{experiment}}\left(  r_{0}\right)  \right]  }\left[  H_{c}^{\left(
\alpha\right)  }\left(  M\sqrt{-V^{\prime\prime}\left(  \phi_{0}\right)
},\left[  M\sqrt{-V^{\prime\prime}\left(  \phi_{0}\right)  }+1\right]
,1,1,\right.  \right. \nonumber\\
&  \left.  \left.  \left[  -4M^{2}V^{\prime\prime}\left(  \phi_{0}\right)
+2M\sqrt{-V^{\prime\prime}\left(  \phi_{0}\right)  }\right]  ;\frac{r_{0}}%
{M}-1\right)  \right]  ^{-1},
\end{align}
where $\phi_{0}=\phi_{\text{cl}}\left(  r_{0}\right)  $.

For $r\rightarrow\infty$, the asymptotics of the quantum fluctuation is
\cite{li2019scattering,li2018scalar,ronveaux1995heun}
\begin{equation}
\eta\sim\frac{M}{r-M}\cos\left(  \sqrt{-V^{\prime\prime}\left(  \phi
_{\text{cl}}\right)  }\left(  r-M\right)  \right.  \left.  +M\sqrt
{-V^{\prime\prime}\left(  \phi_{\text{cl}}\right)  }\ln2\sqrt{-V^{\prime
\prime}\left(  \phi_{\text{cl}}\right)  }\left(  r-M\right)  \right) \nonumber
\end{equation}
and for $r\rightarrow2M$, the asymptotics of the quantum fluctuation is%
\begin{equation}
\eta\sim e^{\sqrt{-V^{\prime\prime}\left(  \phi_{\text{cl}}\right)  }r}\left[
1-\frac{1}{2}\left(  4MV^{\prime\prime}\left(  \phi_{\text{cl}}\right)
+2\sqrt{-V^{\prime\prime}\left(  \phi_{\text{cl}}\right)  }\right)  \left(
r-2M\right)  \right]  .
\end{equation}

\section{$H_{3}$ (Euclidean $AdS_{3}$) spacetime}

The $AdS_{3}$ space and the Euclidean BTZ black hole is important in gravity
\cite{cotler2019theory,lin2019ads,fitzpatrick2016information,dias2019btz,ball2019uplifting}%
. We consider the quantum correction of the gravitational constant in the
three-dimensional hyperbolic space $H_{3}$ (Euclidean anti-de Sitter space
$AdS_{3}$). $H_{3}$ is a three-dimensional subspace of the flat
four-dimensional spacetime: $ds^{2}=dX_{1}^{2}-dT_{1}^{2}+dX_{2}^{2}%
+dT_{2}^{2}$ with the constraint $X_{1}^{2}-T_{1}^{2}+X_{2}^{2}+T_{2}%
^{2}=-l^{2}$ \cite{mann1997quantum,dai2010approach}. The metric is%
\begin{equation}
ds^{2}=l^{2}\left(  d\rho^{2}+\cosh^{2}\rho d\phi^{2}+\sinh^{2}\rho
d\theta^{2}\right)  . \label{metricH3}%
\end{equation}
The equation of the quantum fluctuation (\ref{2.2.6}) with the metric
(\ref{metricH3}) reads%
\begin{align}
&  -\frac{\partial^{2}\eta}{\partial\rho^{2}}-\frac{\cosh^{2}\rho+\sinh
^{2}\rho}{\cosh\rho\sinh\rho}\frac{\partial\eta}{\partial\rho}-\frac{1}%
{\cosh^{2}\rho}\frac{\partial^{2}\eta}{\partial\phi^{2}}-\frac{1}{\sinh
^{2}\rho}\frac{\partial^{2}\eta}{\partial\theta^{2}}\nonumber\\
&  -\left[  l^{2}V^{\prime\prime}\left(  \Phi_{\text{cl}}\right)
+6\chi^{\prime\prime}\left(  \Phi_{\text{cl}}\right)  \right]  \eta=0.
\label{9-4}%
\end{align}
The variable separation $\eta=f\left(  \rho\right)  \cos\left(  k_{1}%
\phi\right)  \cos\left(  k_{2}\theta\right)  $ gives the radial equation of
the quantum fluctuation,%
\begin{align}
&  -\frac{d^{2}f\left(  \rho\right)  }{d\rho^{2}}-\frac{\cosh^{2}\rho
+\sinh^{2}\rho}{\cosh\rho\sinh\rho}\frac{df\left(  \rho\right)  }{d\rho
}\nonumber\\
&  -\left[  \frac{k_{1}^{2}}{\cosh^{2}\rho}+\frac{k_{2}^{2}}{\sinh^{2}\rho
}+l^{2}V^{\prime\prime}\left(  \Phi_{\text{cl}}\right)  +6\chi^{\prime\prime
}\left(  \Phi_{\text{cl}}\right)  \right]  f\left(  \rho\right)  =0.
\label{9-5}%
\end{align}
It is reasonable to assume that the quantum fluctuation is in the ground
state, so we only consider the case of $k_{1}=k_{2}=0$:%
\begin{equation}
\frac{d^{2}f\left(  \rho\right)  }{d\rho^{2}}+\frac{\cosh^{2}\rho+\sinh
^{2}\rho}{\cosh\rho\sinh\rho}\frac{df\left(  \rho\right)  }{d\rho}+\left[
l^{2}V^{\prime\prime}\left(  \Phi_{\text{cl}}\right)  +6\chi^{\prime\prime
}\left(  \Phi_{\text{cl}}\right)  \right]  f\left(  \rho\right)  =0.
\label{9-6}%
\end{equation}

The solution of Eq. (\ref{9-6}) is%
\begin{equation}
f\left(  \rho\right)  =C_{1}\operatorname{P}_{\left[  \sqrt{-l^{2}%
V^{\prime\prime}\left(  \Phi_{\text{cl}}\right)  -6\chi^{\prime\prime}\left(
\Phi_{\text{cl}}\right)  +1}-1\right]  /2}\left(  \cosh2\rho\right)
+C_{2}\operatorname{Q}_{\left[  \sqrt{-l^{2}V^{\prime\prime}\left(
\Phi_{\text{cl}}\right)  -6\chi^{\prime\prime}\left(  \Phi_{\text{cl}}\right)
+1}-1\right]  /2}\left(  \cosh2\rho\right)  ,
\end{equation}
where $\operatorname{P}_{\nu}\left(  z\right)  $ is the Legendre function and
$\operatorname{Q}_{\nu}\left(  z\right)  $ the Legendre function of the second
kind. Since $\operatorname{Q}_{\left[  \sqrt{-l^{2}V^{\prime\prime}\left(
\Phi_{\text{cl}}\right)  -6\chi^{\prime\prime}\left(  \Phi_{\text{cl}}\right)
+1}-1\right]  /2}\left(  \cosh2\rho\right)  $ diverges at $\rho\rightarrow0$,
we choose $C_{2}=0$:%
\begin{equation}
f\left(  \rho\right)  =C\operatorname{P}_{\left[  \sqrt{-l^{2}V^{\prime\prime
}\left(  \Phi_{\text{cl}}\right)  -6\chi^{\prime\prime}\left(  \Phi
_{\text{cl}}\right)  +1}-1\right]  /2}\left(  \cosh2\rho\right)  . \label{9-7}%
\end{equation}
where $C=C_{1}$. Then the quantum fluctuation%
\begin{equation}
\eta=C\operatorname{P}_{\left[  \sqrt{-l^{2}V^{\prime\prime}\left(
\Phi_{\text{cl}}\right)  -6\chi^{\prime\prime}\left(  \Phi_{\text{cl}}\right)
+1}-1\right]  /2}\left(  \cosh2\rho\right)  . \label{9-20}%
\end{equation}

By observation the gravitational constant measured at $\rho_{0}$ is
$G^{\text{experiment}}\left(  \rho_{0}\right)  $. Then $G_{Q}\left(  \rho
_{0}\right)  =G^{\text{experiment}}\left(  \rho_{0}\right)  $, by substituting
Eq. (\ref{9-20}) into (\ref{GQ}), determines the constant $C$:%
\begin{equation}
C=\frac{\sqrt{\frac{2\chi\left(  \Phi_{0}\right)  }{\chi^{\prime\prime}\left(
\Phi_{0}\right)  }\left[  1-16\pi\chi\left(  \Phi_{0}\right)
G^{\text{experiment}}\left(  \rho_{0}\right)  \right]  }}{\operatorname{P}%
_{\left[  \sqrt{-l^{2}V^{\prime\prime}\left(  \Phi_{0}\right)  -6\chi
^{\prime\prime}\left(  \Phi_{0}\right)  +1}-1\right]  /2}\left(  \cosh
2\rho_{0}\right)  },
\end{equation}
where $\Phi_{0}=\Phi_{\text{cl}}\left(  \rho_{0}\right)  $.

\section{$H_{3}/Z$ spacetime}

The $H_{3}/Z$ spacetime, a quotient space of $H_{3}$, is the geometry of the
Euclidean BTZ black hole
\cite{carlip2005conformal,giombi2008one,dai2010approach}. The metric of the
$H_{3}/Z$ (Euclidean $AdS_{3}$) spacetime is given by \cite{giombi2008one}
\begin{equation}
ds^{2}=g_{\mu\nu}dx^{\mu}dx^{\nu}=\frac{dy^{2}+dzd\bar{z}}{y^{2}}%
\end{equation}
with $y=\rho\sin\theta$,$\quad z=\rho\cos\theta e^{i\phi}$ and $1\leq
\rho<e^{2\pi\tau}$, $0\leq\theta<\frac{\pi}{2}$, $0\leq\phi<2\pi$:%
\begin{equation}
ds^{2}=\frac{1}{\rho^{2}\sin^{2}\theta}d\rho^{2}+\frac{1}{\sin^{2}\theta
}d\theta^{2}+\cot^{2}\theta d\phi^{2}. \label{10-1}%
\end{equation}
The equation of the quantum fluctuation is obtained by substituting the metric
(\ref{10-1}) into Eq. (\ref{2.2.6}),
\begin{equation}
-\sin^{2}\theta\left(  \rho^{2}\frac{\partial^{2}\eta}{\partial\rho^{2}}%
+\rho\frac{\partial\eta}{\partial\rho}\right)  -\left(  \sin^{2}\theta
\frac{\partial^{2}\eta}{\partial\theta^{2}}-\tan\theta\frac{\partial\eta
}{\partial\theta}\right)  -\tan^{2}\theta\frac{\partial^{2}\eta}{\partial
\phi^{2}}-\left[  V^{\prime\prime}\left(  \Phi_{\text{cl}}\right)
+6\chi^{\prime\prime}\left(  \Phi_{\text{cl}}\right)  \right]  \eta=0.
\label{10-2}%
\end{equation}
The variable separation $\eta=f\left(  \rho\right)  g\left(  \theta\right)
\cos\left(  k\phi\right)  $ gives%
\begin{align}
&  -\rho^{2}\frac{d^{2}f\left(  \rho\right)  }{d\rho^{2}}-\rho\frac{df\left(
\rho\right)  }{\partial\rho}=\kappa f\left(  \rho\right)  ,\nonumber\\
&  \frac{d^{2}g\left(  \theta\right)  }{d\theta^{2}}-\frac{1}{\sin\theta
\cos\theta}\frac{dg\left(  \theta\right)  }{{d\theta}}+\left\{  \frac{1}%
{\sin^{2}\theta}\left[  V^{\prime\prime}\left(  \Phi_{\text{cl}}\right)
\right.  \right.  \left.  +6\chi^{\prime\prime}\left(  \Phi_{\text{cl}%
}\right)  \right]  \left.  -k^{2}\frac{1}{\cos^{2}\theta}\right\}  g\left(
\theta\right)  =\kappa g\left(  \theta\right)  .
\end{align}
It is reasonable to believe that the world is in its ground state, so we only
consider the ground state of the quantum fluctuation, i.e., $k=0$ and
$\kappa=0$:%
\begin{align}
&  -\rho^{2}\frac{d^{2}f\left(  \rho\right)  }{d\rho^{2}}-\rho\frac{df\left(
\rho\right)  }{\partial\rho}=0,\label{H3Zrho}\\
&  \frac{d^{2}g\left(  \theta\right)  }{d\theta^{2}}-\frac{1}{\sin\theta
\cos\theta}\frac{dg\left(  \theta\right)  }{{d\theta}}+\frac{1}{\sin^{2}%
\theta}\left[  V^{\prime\prime}\left(  \Phi_{\text{cl}}\right)  +6\chi
^{\prime\prime}\left(  \Phi_{\text{cl}}\right)  \right]  g\left(
\theta\right)  =0. \label{H3Ztheta}%
\end{align}

The solution of Eq. (\ref{H3Zrho}) is
\begin{equation}
f\left(  \rho\right)  =C_{1}\ln\rho\label{11-2}%
\end{equation}
and the solution of Eq. (\ref{H3Ztheta}) is
\begin{align}
&  g\left(  \theta\right)  =C_{2}\left(  \sin\theta\right)  ^{-\sqrt
{-V^{\prime\prime}\left(  \Phi_{\text{cl}}\right)  -6\chi^{\prime\prime
}\left(  \Phi_{\text{cl}}\right)  +1}+1}\operatorname{F}\left(  _{-\sqrt
{-V^{\prime\prime}\left(  \Phi_{\text{cl}}\right)  -6\chi^{\prime\prime
}\left(  \Phi_{\text{cl}}\right)  +1}+1}^{^{\left[  -\sqrt{-V^{\prime\prime
}\left(  \Phi_{\text{cl}}\right)  -6\chi^{\prime\prime}\left(  \Phi
_{\text{cl}}\right)  +1}+1\right]  /2,\left[  -\sqrt{-V^{\prime\prime}\left(
\Phi_{\text{cl}}\right)  -6\chi^{\prime\prime}\left(  \Phi_{\text{cl}}\right)
+1}+1\right]  /2}};\sin^{2}\theta\right) \nonumber\\
&  +C_{3}\left(  \sin\theta\right)  ^{\sqrt{-V^{\prime\prime}\left(
\Phi_{\text{cl}}\right)  -6\chi^{\prime\prime}\left(  \Phi_{\text{cl}}\right)
+1}+1}\operatorname{F}\left(  _{\sqrt{-V^{\prime\prime}\left(  \Phi
_{\text{cl}}\right)  -6\chi^{\prime\prime}\left(  \Phi_{\text{cl}}\right)
+1}+1}^{\left[  \sqrt{-V^{\prime\prime}\left(  \Phi_{\text{cl}}\right)
-6\chi^{\prime\prime}\left(  \Phi_{\text{cl}}\right)  +1}+1\right]  /2,\left[
\sqrt{-V^{\prime\prime}\left(  \Phi_{\text{cl}}\right)  -6\chi^{\prime\prime
}\left(  \Phi_{\text{cl}}\right)  +1}+1\right]  /2};\sin^{2}\theta\right)  ,
\label{gtheta}%
\end{align}
where $\operatorname{F}\left(  _{c}^{a,b};z\right)  =_{2}\operatorname{F}%
_{1}\left(  a;b;c,z\right)  $ is the hypergeometric function
\cite{beyer2018handbook}. The first term in Eq. (\ref{gtheta}) diverges when
$\theta=0$, so we choose $C_{2}=0$ and then the solution reads
\begin{equation}
g\left(  \theta\right)  =C_{3}\left(  \sin\theta\right)  ^{\sqrt
{-V^{\prime\prime}\left(  \Phi_{\text{cl}}\right)  -6\chi^{\prime\prime
}\left(  \Phi_{\text{cl}}\right)  +1}+1}\operatorname{F}\left(  _{\sqrt
{-V^{\prime\prime}\left(  \Phi_{\text{cl}}\right)  -6\chi^{\prime\prime
}\left(  \Phi_{\text{cl}}\right)  +1}+1}^{\left[  \sqrt{-V^{\prime\prime
}\left(  \Phi_{\text{cl}}\right)  -6\chi^{\prime\prime}\left(  \Phi
_{\text{cl}}\right)  +1}+1\right]  /2,\left[  \sqrt{-V^{\prime\prime}\left(
\Phi_{\text{cl}}\right)  -6\chi^{\prime\prime}\left(  \Phi_{\text{cl}}\right)
+1}+1\right]  /2};\sin^{2}\theta\right)  .
\end{equation}

Then the quantum fluctuation $\eta=f\left(  \rho\right)  g\left(
\theta\right)  $ is
\begin{equation}
\eta=C\ln\rho\left(  \sin\theta\right)  ^{\sqrt{-V^{\prime\prime}\left(
\Phi_{\text{cl}}\right)  -6\chi^{\prime\prime}\left(  \Phi_{\text{cl}}\right)
+1}+1}\operatorname{F}\left(  _{\sqrt{-V^{\prime\prime}\left(  \Phi
_{\text{cl}}\right)  -6\chi^{\prime\prime}\left(  \Phi_{\text{cl}}\right)
+1}+1}^{\left[  \sqrt{-V^{\prime\prime}\left(  \Phi_{\text{cl}}\right)
-6\chi^{\prime\prime}\left(  \Phi_{\text{cl}}\right)  +1}+1\right]  /2,\left[
\sqrt{-V^{\prime\prime}\left(  \Phi_{\text{cl}}\right)  -6\chi^{\prime\prime
}\left(  \Phi_{\text{cl}}\right)  +1}+1\right]  /2};\sin^{2}\theta\right)  .
\label{11-12}%
\end{equation}

By observation the gravitational constant measured at $\left(  \rho_{0}%
,\theta_{0}\right)  $ is $G^{\text{experiment}}\left(  \rho_{0},\theta
_{0}\right)  $. Then $G_{Q}\left(  \rho_{0},\theta_{0}\right)
=G^{\text{experiment}}\left(  \rho_{0},\theta_{0}\right)  $, by Eq.
(\ref{GQ}), determines the constant $C$:%
\begin{align}
C  &  =\sqrt{\frac{2\chi\left(  \Phi_{0}\right)  }{\chi^{\prime\prime}\left(
\Phi_{0}\right)  }\left[  1-16\pi\chi\left(  \Phi_{0}\right)
G^{\text{experiment}}\left(  \rho_{0},\theta_{0}\right)  \right]  }\left[
\ln\rho_{0}\left(  \sin\theta_{0}\right)  ^{\sqrt{-V^{\prime\prime}\left(
\Phi_{0}\right)  -6\chi^{\prime\prime}\left(  \Phi_{0}\right)  +1}+1}\right.
\nonumber\\
&  \times\left.  \operatorname{F}\left(  _{\sqrt{-V^{\prime\prime}\left(
\Phi_{0}\right)  -6\chi^{\prime\prime}\left(  \Phi_{0}\right)  +1}+1}^{\left[
\sqrt{-V^{\prime\prime}\left(  \Phi_{0}\right)  -6\chi^{\prime\prime}\left(
\Phi_{0}\right)  +1}+1\right]  /2,\left[  \sqrt{-V^{\prime\prime}\left(
\Phi_{0}\right)  -6\chi^{\prime\prime}\left(  \Phi_{0}\right)  +1}+1\right]
/2};\sin^{2}\theta_{0}\right)  \right]  ^{-1},
\end{align}
where $\Phi_{0}=\Phi_{\text{cl}}\left(  \rho_{0},\theta_{0}\right)  $.

\section{Universe model}

Next we consider the quantum correction of the gravitational constant in\ the
universe model.

In the spacetime described by the Robertson-Walker metric,%
\begin{equation}
g_{\mu\nu}=\operatorname*{diag}\left(  -1,\frac{a^{2}\left(  t\right)
}{1-kr^{2}},a^{2}\left(  t\right)  r^{2},a^{2}\left(  t\right)  r^{2}\sin
^{2}\theta\right)  \label{4.1.1}%
\end{equation}
with $a\left(  t\right)  $ the scale factor and $k$ the curvature, the field
equation of the scalar field $\Phi$ by the action (\ref{2.1.0}) reads%
\begin{equation}
\ddot{\Phi}+3\frac{\dot{a}}{a}\dot{\Phi}-V^{\prime}\left(  \Phi\right)
+6\chi^{\prime}\left(  \Phi\right)  \left(  \frac{\ddot{a}}{a}+\frac{\dot
{a}^{2}}{a^{2}}+\frac{k}{a^{2}}\right)  =0. \label{4.1.2}%
\end{equation}
Taking only the leading contribution into account, we here use the classical
solution of the scale factor $a\left(  t\right)  $.

The equation of the quantum fluctuation by Eq. (\ref{2.2.6}) reads%
\begin{equation}
\ddot{\eta}+3\frac{\dot{a}}{a}\dot{\eta}-\left[  V^{\prime\prime}\left(
\Phi_{\text{cl}}\right)  -6\chi^{\prime\prime}\left(  \Phi_{\text{cl}}\right)
\left(  \frac{\ddot{a}}{a}+\frac{\dot{a}^{2}}{a^{2}}+\frac{k}{a^{2}}\right)
\right]  \eta=0. \label{4.1.3}%
\end{equation}
The scale factor $a$ is given by the universe model.

\subsection{Friedmann universe}

For the Friedmann universe with the curvature $k=0$, when the energy density
of the universe is dominated by the non-relativistic matter, the scale factor
$a\left(  t\right)  $ is \cite{ohanian2013gravitation}%
\begin{equation}
a\left(  t\right)  =\left(  \frac{3GM}{\pi}\right)  ^{1/3}t^{2/3},
\label{4.2.1}%
\end{equation}
where $M=2\pi^{2}a^{3}\rho$ is the mass of the universe with $\rho$ the
density. With the scale factor given by Eq. (\ref{4.2.1}), the equation of the
quantum fluctuation reads%
\begin{equation}
\ddot{\eta}+\frac{2}{t}\dot{\eta}-\left[  V^{\prime\prime}\left(
\Phi_{\text{cl}}\right)  -\frac{4\chi^{\prime\prime}\left(  \Phi_{\text{cl}%
}\right)  }{3t^{2}}\right]  \eta=0. \label{4.2.3}%
\end{equation}
Solving Eq. (\ref{4.2.3}) gives%
\begin{equation}
\eta\left(  t\right)  =C_{1}\frac{1}{\sqrt{t}}I_{\sqrt{\frac{1}{4}-\frac{4}%
{3}\chi^{\prime\prime}\left(  \Phi_{\text{cl}}\right)  }}\left(
\sqrt{V^{\prime\prime}\left(  \Phi_{\text{cl}}\right)  }t\right)  +C_{2}%
\frac{1}{\sqrt{t}}K_{^{\sqrt{\frac{1}{4}-\frac{4}{3}\chi^{\prime\prime}\left(
\Phi_{\text{cl}}\right)  }}}\left(  \sqrt{V^{\prime\prime}\left(
\Phi_{\text{cl}}\right)  }t\right)  ,
\end{equation}
where $I_{\nu}\left(  z\right)  $ is the Modified Bessel function of the first
kind and $K_{\nu}\left(  z\right)  $ is the Modified Bessel function of the
second kind. The Bessel functions $I_{\nu}\left(  z\right)  $ and $K_{\nu
}\left(  z\right)  $ have the asymptotics $I_{\nu}\left(  z\right)
\sim\left(  z/2\right)  ^{\nu}$ and $K_{\nu}\left(  z\right)  \sim\left(
z/2\right)  ^{-\nu}$. Then for $t\rightarrow0$ we have the asymptotics
$\eta\left(  t\right)  \sim C_{1}\left(  \frac{\sqrt{V^{\prime\prime}\left(
\Phi_{\text{cl}}\right)  }}{2}\right)  ^{\sqrt{\frac{1}{4}-\frac{4}{3}%
\chi^{\prime\prime}\left(  \Phi_{\text{cl}}\right)  }}$ $t^{\sqrt{\frac{1}%
{4}-\frac{4}{3}\chi^{\prime\prime}\left(  \Phi_{\text{cl}}\right)  }-\frac
{1}{2}}$ $+C_{2}\left(  \frac{\sqrt{V^{\prime\prime}\left(  \Phi_{\text{cl}%
}\right)  }}{2}\right)  ^{\sqrt{\frac{1}{4}-\frac{4}{3}\chi^{\prime\prime
}\left(  \Phi_{\text{cl}}\right)  }}$ $t^{-\sqrt{\frac{1}{4}-\frac{4}{3}%
\chi^{\prime\prime}\left(  \Phi_{\text{cl}}\right)  }-\frac{1}{2}}$. Here
$\sqrt{\frac{1}{4}-\frac{4}{3}\chi^{\prime\prime}\left(  \Phi_{\text{cl}%
}\right)  }>\frac{1}{2}$ for $\chi^{\prime\prime}\left(  \Phi_{\text{cl}%
}\right)  <0$, so the $K_{\nu}\left(  z\right)  $ term diverges at
$t\rightarrow0$. This requires $C_{2}=0$. Then quantum fluctuation reads%
\begin{equation}
\eta\left(  t\right)  =C\frac{1}{\sqrt{t}}I_{\sqrt{\frac{1}{4}-\frac{4}{3}%
\chi^{\prime\prime}\left(  \Phi_{\text{cl}}\right)  }}\left(  \sqrt
{V^{\prime\prime}\left(  \Phi_{\text{cl}}\right)  }t\right)  .
\end{equation}

By observation the gravitational constant measured at $t_{0}$ is
$G^{\text{experiment}}\left(  t_{0}\right)  $. Then $G_{Q}\left(
t_{0}\right)  =G^{\text{experiment}}\left(  t_{0}\right)  $, by Eq.
(\ref{GQ}), determines the constant $C$:%
\begin{equation}
C=\frac{\sqrt{\frac{2\chi\left(  \Phi_{0}\right)  }{\chi^{\prime\prime}\left(
\Phi_{0}\right)  }\left[  1-16\pi\chi\left(  \Phi_{0}\right)
G^{\text{experiment}}\left(  t_{0}\right)  \right]  }}{\frac{1}{\sqrt{t_{0}}%
}I_{\sqrt{\frac{1}{4}-\frac{4}{3}\chi^{\prime\prime}\left(  \Phi_{0}\right)
}}\left(  \sqrt{V^{\prime\prime}\left(  \Phi_{0}\right)  }t_{0}\right)  },
\end{equation}
where $\Phi_{0}=\Phi_{\text{cl}}\left(  t_{0}\right)  $.

\subsection{Inflationary universe}

In the inflationary universe epoch the scale factor is
\cite{ohanian2013gravitation}
\begin{equation}
a\left(  t\right)  =\exp\left(  \sqrt{\frac{8\pi G\rho_{v}}{3}}t\right)  ,
\label{4.3.1}%
\end{equation}
where $\rho_{v}$ is the vacuum energy density and the curvature $k=0$. Then by
Eq. (\ref{4.1.3}) the equation of the quantum fluctuation in the inflationary
epoch reads%

\begin{equation}
\ddot{\eta}+3\sqrt{\frac{8\pi G\rho_{v}}{3}}\dot{\eta}-\left[  V^{\prime
\prime}\left(  \Phi_{\text{cl}}\right)  -32\pi G\rho_{v}\chi^{\prime\prime
}\left(  \Phi_{\text{cl}}\right)  \right]  \eta=0. \label{4.3.2}%
\end{equation}
In\ the inflationary epoch, we concern ourselves with the epoch $0<t<t_{0}$:
\begin{equation}
\eta=C\exp\left(  -\sqrt{6\pi G\rho_{v}}\left[  1+\sqrt{1-\frac{1}{6\pi
G\rho_{v}}V^{\prime\prime}\left(  \Phi_{\text{cl}}\right)  -\frac{16}{3}%
\chi^{\prime\prime}\left(  \Phi_{\text{cl}}\right)  }\right]  t\right)  .
\label{4.3.3}%
\end{equation}

By observation the gravitational constant measured at $t_{0}$ is
$G^{\text{experiment}}\left(  t_{0}\right)  $. Then $G_{Q}\left(
t_{0}\right)  =G^{\text{experiment}}\left(  t_{0}\right)  $, by Eq.
(\ref{GQ}), determines the constant $C$:%
\begin{equation}
C=\frac{\sqrt{\frac{2\chi\left(  \Phi_{0}\right)  }{\chi^{\prime\prime}\left(
\Phi_{0}\right)  }\left[  1-16\pi\chi\left(  \Phi_{0}\right)
G^{\text{experiment}}\left(  t_{0}\right)  \right]  }}{\exp\left(  -\sqrt{6\pi
G\rho_{v}}\left[  1+\sqrt{1-\frac{1}{6\pi G\rho_{v}}V^{\prime\prime}\left(
\Phi_{0}\right)  -\frac{16}{3}\chi^{\prime\prime}\left(  \Phi_{0}\right)
}\right]  t_{0}\right)  },
\end{equation}
where $\Phi_{0}=\Phi_{\text{cl}}\left(  t_{0}\right)  $.

\section{de Sitter spacetime}

The de Sitter spacetime is described by the Robertson-Walker metric with
\begin{equation}
a\left(  t\right)  =a_{0}e^{\sqrt{\frac{\Lambda}{3}}t} \label{dSat}%
\end{equation}
and the curvature $k=0$, where $\Lambda$ is the cosmological constant
\cite{ohanian2013gravitation}. The equation of the field $\Phi$ by the action%
\begin{equation}
I=\int d^{4}x\sqrt{-g}\left[  \chi\left(  \Phi\right)  \left(  R-2\Lambda
\right)  +\frac{1}{2}g^{\mu\nu}\partial_{\mu}\Phi\partial_{\nu}\Phi-V\left(
\Phi\right)  \right]  \label{7.1.1}%
\end{equation}
reads%
\begin{equation}
\ddot{\Phi}+3\frac{\dot{a}}{a}\dot{\Phi}-V^{\prime}\left(  \Phi\right)
+6\left(  \frac{\ddot{a}}{a}+\frac{\dot{a}^{2}}{a^{2}}-\frac{\Lambda}%
{3}\right)  \chi^{\prime}\left(  \Phi\right)  =0. \label{7.1.3}%
\end{equation}

The equation of the quantum fluctuation by substituting $\Phi=\Phi_{\text{cl}%
}+\eta$ into Eq. (\ref{7.1.3}), up to the leading order, is%
\begin{equation}
\ddot{\eta}+3\frac{\dot{a}}{a}\dot{\eta}-\left[  V^{\prime\prime}\left(
\Phi_{\text{cl}}\right)  -6\chi^{\prime\prime}\left(  \Phi_{\text{cl}}\right)
\left(  \frac{\ddot{a}}{a}+\frac{\dot{a}^{2}}{a^{2}}-\frac{\Lambda}{3}\right)
\right]  \eta=0. \label{7.1.4}%
\end{equation}
For $\Lambda>0$, by Eq. (\ref{dSat}), we have%
\begin{equation}
\ddot{\eta}+3\sqrt{\frac{\Lambda}{3}}\dot{\eta}+\left[  2\Lambda\chi
^{\prime\prime}\left(  \Phi_{\text{cl}}\right)  -V^{\prime\prime}\left(
\Phi_{\text{cl}}\right)  \right]  \eta=0. \label{7.1.5}%
\end{equation}
The solution is%
\begin{equation}
\eta=C\exp\left(  -\frac{1}{2}\left[  \sqrt{3\Lambda}+\sqrt{3\Lambda
-8\Lambda\chi^{\prime\prime}\left(  \Phi_{\text{cl}}\right)  +4V^{\prime
\prime}\left(  \Phi_{\text{cl}}\right)  }\right]  t\right)  . \label{7.1.6}%
\end{equation}

By observation the gravitational constant measured at $t_{0}$ is
$G^{\text{experiment}}\left(  t_{0}\right)  $. Then $G_{Q}\left(
t_{0}\right)  =G^{\text{experiment}}\left(  t_{0}\right)  $, by Eq.
(\ref{GQ}), determines the constant $C$:%
\begin{equation}
C=\frac{\sqrt{\frac{2\chi\left(  \Phi_{0}\right)  }{\chi^{\prime\prime}\left(
\Phi_{0}\right)  }\left[  1-16\pi\chi\left(  \Phi_{0}\right)
G^{\text{experiment}}\left(  t_{0}\right)  \right]  }}{\exp\left(  -\frac
{1}{2}\left[  \sqrt{3\Lambda}+\sqrt{3\Lambda-8\Lambda\chi^{\prime\prime
}\left(  \Phi_{0}\right)  +4V^{\prime\prime}\left(  \Phi_{0}\right)  }\right]
t_{0}\right)  },
\end{equation}
where $\Phi_{0}=\Phi_{\text{cl}}\left(  t_{0}\right)  $.

\section{Rindler spacetime}

The Rindler spacetime is a spacetime region that a uniformly accelerated
observer sees. For a uniformly accelerated observer with the acceleration $a$
in the Minkowski spacetime \cite{Misner1974Gravitation,Rindler1977Essential},
$ds^{2}=-dT^{2}+dX^{2}+dY^{2}+dZ^{2}$, the Rindler spacetime is described by
the metric
\begin{equation}
ds^{2}=\pm e^{2a\xi}\left(  -d\tau^{2}+d\xi^{2}\right)  +dY^{2}+dZ^{2},
\label{6.1.1}%
\end{equation}
where "$+$" corresponds to the R and L regions and "$-$" corresponds to the F
and P regions in the Rindler spacetime. In the R region, $T=\frac{1}{a}%
e^{a\xi}\sinh\left(  a\tau\right)  $ and $X=\frac{1}{a}e^{a\xi}\cosh\left(
a\tau\right)  $, and in the L region, $T=-\frac{1}{a}e^{a\xi}\sinh\left(
a\tau\right)  $ and $X=-\frac{1}{a}e^{a\xi}\cosh\left(  a\tau\right)  $. In
the R and L regions, $\tau$ is the time and $\xi$ is the spatial coordinate
with $\left\vert T\right\vert \leq\left\vert X\right\vert $. In the F region
$T=\frac{1}{a}e^{a\xi}\cosh\left(  a\tau\right)  $ and $X=\frac{1}{a}e^{a\xi
}\sinh\left(  a\tau\right)  $, and in the P region $T=-\frac{1}{a}e^{a\xi
}\cosh\left(  a\tau\right)  $ and $X=-\frac{1}{a}e^{a\xi}\sinh\left(
a\tau\right)  $. In the F and P regions, $\xi$ is the time and $\tau$ is the
spatial coordinate with $\left\vert T\right\vert \geq\left\vert X\right\vert
$. Here the Rindler coordinates $\left(  \tau,\xi\right)  $ are the comoving
coordinates of the uniformly accelerated observer along the $X$ axis:
\begin{align}
\tau &  =\frac{1}{a}\operatorname{arctanh}\frac{T}{X},\nonumber\\
\xi &  =\frac{1}{a}\ln\left(  a\sqrt{X^{2}-T^{2}}\right)  .
\end{align}

The equation of the unobservable field $\Phi$, by Eq. (\ref{2.1.0}), reads%
\begin{equation}
\mp e^{-2a\xi}\frac{\partial^{2}\Phi}{\partial\tau^{2}}\pm e^{-2a\xi}%
\frac{\partial^{2}\Phi}{\partial\xi^{2}}+\frac{\partial^{2}\Phi}{\partial
Y^{2}}+\frac{\partial^{2}\Phi}{\partial Z^{2}}+V^{\prime}\left(  \Phi\right)
=0. \label{6.1.2}%
\end{equation}

In the classical picture, the gravitational constant given by Eq.
(\ref{2.1.9}) is a constant and is related to the classical ground-state
solution which is a constant solution. If the unobservable field $\Phi$ is a
quantum field, then there is a quantum correction to the gravitational
constant. Writing the quantum unobservable field as $\Phi=\Phi_{\text{cl}%
}+\eta$ and substituting into Eq. (\ref{6.1.2}), after variable separation
$\eta=\cos\left(  \omega t\right)  \phi\left(  \xi\right)  P\left(
Y,Z\right)  $, we arrive at a stationary equation of $\phi\left(  \xi\right)
$,%
\begin{equation}
\pm\frac{\partial^{2}\phi\left(  \xi\right)  }{\partial\xi^{2}}+\left[  \pm
e^{-2a\xi}\omega^{2}+V^{\prime\prime}\left(  \Phi_{\text{cl}}\right)
-\kappa_{1}^{2}-\kappa_{2}^{2}\right]  e^{2a\xi}\phi\left(  \xi\right)  =0
\label{6.1.5}%
\end{equation}
and $P\left(  Y,Z\right)  =\cos\left(  \kappa_{1}Y\right)  \cos\left(
\kappa_{2}Z\right)  $, where $\kappa_{1}$ and $\kappa_{2}$ are\ the momenta in
the $Y$ and $Z$ directions.

In the classical picture the gravitational constant corresponds to the
classical ground state, so we only need to consider the quantum correction of
the ground state. For the ground state, $\omega=0$, $\kappa_{1}=\kappa_{2}=0$
and then $\eta=\phi\left(  \xi\right)  $. Eq. (\ref{6.1.5}) becomes%

\begin{equation}
\pm\frac{\partial^{2}\phi\left(  \xi\right)  }{\partial\xi^{2}}+V^{\prime
\prime}\left(  \Phi_{\text{cl}}\right)  e^{2a\xi}\phi\left(  \xi\right)  =0.
\label{6.1.6}%
\end{equation}

In the R and L regions, Eq. (\ref{6.1.6}) becomes
\begin{equation}
\frac{\partial^{2}\phi\left(  \xi\right)  }{\partial\xi^{2}}-V^{\prime\prime
}\left(  \Phi_{\text{cl}}\right)  e^{2a\xi}\phi\left(  \xi\right)  =0.
\end{equation}
The solution is%
\begin{equation}
\phi\left(  \xi\right)  =C_{1}K_{0}\left(  \sqrt{V^{\prime\prime}\left(
\Phi_{cl}\right)  }\frac{e^{a\xi}}{a}\right)  +C_{1}^{\prime}I_{0}\left(
\sqrt{V^{\prime\prime}\left(  \Phi_{cl}\right)  }\frac{e^{a\xi}}{a}\right)  ,
\end{equation}
where $I_{\nu}\left(  z\right)  $ is the modified Bessel function of the first
kind and $K_{\nu}\left(  z\right)  $ is modified Bessel function of the second
kind. Requiring that $\phi\left(  \xi\right)  $ is finite at $\xi
\rightarrow\infty$, we have%
\begin{equation}
\eta=\phi\left(  \xi\right)  =CK_{0}\left(  \sqrt{V^{\prime\prime}\left(
\Phi_{cl}\right)  }\frac{e^{a\xi}}{a}\right)  .
\end{equation}
where $C=C_{1}$. Then the quantum correction to the gravitational constant, by
Eqs. (\ref{XiPhi}) and (\ref{GQ}), is
\begin{equation}
G_{Q}=\frac{1}{16\pi\chi\left(  \Phi_{\text{cl}}\right)  }\left[
1-\frac{C^{2}}{2}\frac{\chi^{\prime\prime}\left(  \Phi_{cl}\right)  }%
{\chi\left(  \Phi_{cl}\right)  }K_{0}^{2}\left(  \sqrt{V^{\prime\prime}\left(
\Phi_{cl}\right)  }\frac{e^{a\xi}}{a}\right)  \right]  . \label{6.1.12}%
\end{equation}

By observation the gravitational constant measured at $\xi_{0}$ is
$G^{\text{experiment}}\left(  \xi_{0}\right)  $. Then $G_{Q}\left(  \xi
_{0}\right)  =G^{\text{experiment}}\left(  \xi_{0}\right)  $ determines the
constant $C$:%
\begin{equation}
C=\frac{\sqrt{\frac{2\chi\left(  \Phi_{0}\right)  }{\chi^{\prime\prime}\left(
\Phi_{0}\right)  }\left[  1-16\pi\chi\left(  \Phi_{0}\right)
G^{\text{experiment}}\left(  \xi_{0}\right)  \right]  }}{K_{0}\left(
\sqrt{V^{\prime\prime}\left(  \Phi_{0}\right)  }\frac{e^{a\xi_{0}}}{a}\right)
},
\end{equation}
where $\Phi_{0}=\Phi_{\text{cl}}\left(  \xi_{0}\right)  $.

In the F and P regions, Eq. (\ref{6.1.6}) becomes%
\begin{equation}
\frac{\partial^{2}\phi\left(  \xi\right)  }{\partial\xi^{2}}+V^{\prime\prime
}\left(  \Phi_{\text{cl}}\right)  e^{2a\xi}\phi\left(  \xi\right)  =0.
\end{equation}
The solution is%
\begin{equation}
\phi\left(  \xi\right)  =C_{2}J_{0}\left(  \sqrt{V^{\prime\prime}\left(
\Phi_{\text{cl}}\right)  }\frac{e^{a\xi}}{a}\right)  +C_{2}^{\prime}%
Y_{0}\left(  \sqrt{V^{\prime\prime}\left(  \Phi_{\text{cl}}\right)  }%
\frac{e^{a\xi}}{a}\right)  . \label{6.1.10}%
\end{equation}
where $J_{\nu}\left(  z\right)  $ is the Bessel function of the first kind and
$Y_{\nu}\left(  z\right)  $ is the Bessel function of the second kind. \ 

Dropping the $Y_{0}$ term which diverges at $\xi\rightarrow0$ when $a$ is
large, we have%
\begin{equation}
\eta=\phi\left(  \xi\right)  =CJ_{0}\left(  \sqrt{V^{\prime\prime}\left(
\Phi_{\text{cl}}\right)  }\frac{e^{a\xi}}{a}\right)  , \label{6.1.11}%
\end{equation}
where $C=C_{2}$.

Then the quantum correction of the gravitational constant, by Eqs.
(\ref{XiPhi}) and (\ref{GQ}), is
\begin{equation}
G_{Q}=\frac{1}{16\pi\chi\left(  \Phi_{\text{cl}}\right)  }\left[
1-\frac{C^{2}}{2}\frac{\chi^{\prime\prime}\left(  \Phi_{\text{cl}}\right)
}{\chi\left(  \Phi_{\text{cl}}\right)  }J_{0}^{2}\left(  \sqrt{V^{\prime
\prime}\left(  \Phi_{\text{cl}}\right)  }\frac{e^{a\xi}}{a}\right)  \right]  .
\end{equation}

By observation the gravitational constant measured at $\xi_{0}$ is
$G^{\text{experiment}}\left(  \xi_{0}\right)  $. Then $G_{Q}\left(  \xi
_{0}\right)  =G^{\text{experiment}}\left(  \xi_{0}\right)  $ determines the
constant $C$:%
\begin{equation}
C=\frac{\sqrt{\frac{2\chi\left(  \Phi_{0}\right)  }{\chi^{\prime\prime}\left(
\Phi_{0}\right)  }\left[  1-16\pi\chi\left(  \Phi_{0}\right)
G^{\text{experiment}}\left(  \xi_{0}\right)  \right]  }}{J_{0}\left(
\sqrt{V^{\prime\prime}\left(  \Phi_{0}\right)  }\frac{e^{a\xi_{0}}}{a}\right)
},
\end{equation}
where $\Phi_{0}=\Phi_{\text{cl}}\left(  \xi_{0}\right)  $.

\section{Conclusions and outlook \label{Conclusion}}

In this paper, we suggest a scheme for considering the quantum correction of
the gravitational constant. In the model, the gravitational constant is
induced by the coupling of the gravitational field and a scalar field. When
the scalar field is a quantum field, there will be a time- and space-dependent
quantum correction of the gravitational constant. The order of magnitude of
the quantum correction is of the magnitude of Planck's constant.

The focus of this paper is to give a conceptual and qualitative discussion on
the quantum correction of the gravitational constant. In order to illustrate
the main concept clearly, we construct the toy models as simply as possible.
In a more realistic and therefore more complex case, the choice of the scalar
field will take into account more factors, such as symmetry breaking. But the
qualitative conclusion remains the same. When needed, along this line of
thought, such a consideration can be applied to various models.

Moreover, the quantum-corrected gravitational constant varies with time and
space, which will influence the black hole and the Hawking radiation. In
further work, we will consider the black hole solution under the
quantum-corrected gravitational constant and its influence on the Hawking
radiation, especially the species problem of the black hole entropy
\cite{chen2018entropy}. The gravitational wave with a quantum corrected
gravitational constant is also worth considering \cite{li2021gravitational}.
The gravitational constant in the model originates from a coupling between
gravity and a scalar field. In principle we need to consider the solution of a
scalar field in a certain spacetime. Previously, the solution of a scalar
field in a spacetime is solved under the premise that the gravitational
constant is a constant
\cite{doran2002perturbation,futterman1987scattering,pike2002scattering,macedo2014absorption}%
. The quantum-corrected gravitational constant, however, is no longer a
constant, so we need to reconsider the solution of the scalar field in curved
spacetime under the time- and space-dependent quantum-corrected gravitational
constant. The measurement of the gravitational constant is always a basic
experimental problem in physics
\cite{xue2020precision,rosi2014precision,quinn2000measuring}.

\

\bigskip

\acknowledgments

We are very indebted to Dr G. Zeitrauman for his encouragement. This work is supported in part by Special Funds for theoretical physics Research Program of the NSFC under Grant No. 11947124, Nankai Zhide foundation, and NSF of China under Grant No. 11575125 and No. 11675119.

%(正文结束)——————————————————————————————————————————————————

%%%%%%%%%%%%%%%%%%%参考文献%%%%%%%%%%%%%%%%%%%%%%%%%%%

%%%%%%%%%%%%%%%%%%%bibtex形式的参考文献%%%%%%%%%%%%%%%
%\bibliographystyle{JHEP} %参考文献的风格(.bst)
%\bibliography{refs} %参考文献文件(.bib)
%\nocite{*} %若去掉注释，没有被引用的文献也被列出

%%%%%%%%%%%%%%%%%%%bbl形式的参考文献%%%%%%%%%%%%%%%%%%

%%%%%%%%%%%%%%%%%%%%%%%%%%%%%%%%%%%%%%%%%%%%%%%%%%%%%%

\end{document}